\documentstyle[prl,eqsecnum,aps]{revtex}

\begin{document}
\draft
\author{Jian-Qi Shen, Hong-Yi Zhu, and Shen-Lei Shi}
\title{Gravitomagnetic Field and Time-Dependent Spin-Rotation Coupling }
\address{1.State Key Laboratory of Modern Optical Instrumentation, Center for Optical%
\\
and Electromagnetic Research, College of Information Science and Engineering%
\\
2.Zhejiang Institute of Modern Physics and Department of Physics,\\
Zhejiang University, Hangzhou 310027, People$^{,}$s Republic of China}
\date{\today }
\maketitle

\begin{abstract}
The Kerr metric of spherically symmetric gravitational field is analyzed
through the coordinate transformation from the rotating frame to fixing
frame, and consequently that the inertial force field (with the exception of
the centrifugal force field) in the rotating system is one part of its
gravitomagnetic field is verified. We investigate the spin-rotation coupling
and, by making use of Lewis-Riesenfeld invariant theory, we obtain exact
solutions of the Schr\"{o}dinger equation of a spinning particle in a
time-dependent rotating reference frame. A potential application of these
exact solutions to the investigation of Earth$^{,}$s rotating frequency
fluctuation by means of neutron-gravity interferometry experiment is briefly
discussed in the present paper.
\end{abstract}


\section{Introduction}

One can easily verify that the field equation of general relativity in
low-motion weak-field approximation is somewhat analogous to Maxwell$^{,}$s
equation of electromagnetic field. It is the most outstanding point that the
former field (gravitational field) also possesses both the gravitoelectric
potential written as $\frac{g_{00}-1}{2}$ and the gravitomagnetic potentials
as $\vec{A}=(g_{01},g_{02},g_{03}),$ and the corresponding gravitomagnetic
field strength is of the form $\vec{B}=-\frac{1}{2}\nabla \times \vec{A}$. A
particle with intrinsic spin possesses a gravitomagnetic moment of such
magnitude that it equals the spin of this particle. The interaction between
the gravitomagnetic moment and the gravitomagnetic field is thus also called
the spin-gravity coupling\cite{Kleinert,Mashhoon1}, of which the Hamiltonian
is given by

\begin{equation}
H=\frac{1}{2}\vec{B}\cdot \vec{S}.  \label{eq1}
\end{equation}

\ It is shown in what follows that the strongest gravitomagnetic field that
we can find on the Earth arises from the Earth$^{,}$s rotation, that is, the
Earth$^{,}$s rotation gives rise to an inertial gravitomagnetic force field
observed in the rotating frame. Since the Earth is a noninertial reference
frame due to its rotation, a spinning particle is coupled to a more strong
gravitomagnetic field (i.e., Earth$^{,}$s rotation frequency), which
represents the coupling of spin-noninertial frame in addition to the
interaction expressed by Eq. (\ref{eq1}). It is apparently seen that the
interaction of angular momentum of a particle with noninertial frame is
related to the Coriolis force\cite{Shen}. These two gravitomagnetic fields
(see in Sec.2 for detailed differences between them) have different origins
and properties: the gravitomagnetic field caused by mass current, expressed
by $\vec{B}=-\frac{1}{2}\nabla \times \vec{A},$ is similar to the magnetic
field produced by electric current, and its strength is dependent on the
Newtonian gravitational constant $G,$ while the gravitomagnetic field
associated with the Coriolis force depends on the choice of the coordinates
and in consequence its strength is independent of the Newtonian
gravitational constant. That is, in accordance with Newton$^{,}$s law the
coordinate transformation from the rotating frame to the fixing frame
results in this inertial force observed by the observer fixed in the
rotating reference frame. Apparently, due to the smallness of $G,$ the
coupling of the latter gravitomagnetic field with intrinsic spin is $10^{20}$
times stronger than that of Eq. (\ref{eq1})\cite{Mashhoon2}.

In the present paper, we further investigate the interaction between this
inertial force field and the intrinsic spin of a particle. According to the
equivalence principle, the nature of the inertial force is gravitational
force, and consequently both expressions of these two gravitomagnetic forces
(namely, the gravitational Lorentz force and Coriolis force) can be derived
from the equation of gravitational field. This work is given in what follows
and we thus obtain the Hamiltonian of the spin-rotation coupling. It is
known that Mashhoon$^{,}$s approach to deriving the intrinsic spin -rotation
coupling is suggested by analyzing the Doppler$^{,}$s effect of wavelight in
the rotating frame with respect to the fixing frame \cite
{Mashhoon1,Mashhoon2}. In this paper, however, the transformation of the
gravitomagnetic potentials is studied through the coordinate transformation,
and as a result, the Hamiltonian of the coupling of the intrinsic spin of a
particle with the rotating frequency of a rotating reference frame is then
obtained.

The reason why the coupling of spin (or gravitomagnetic moment) with
noninertial frame is of great importance lies in that, with the development
of laser technology and their applications to the gravitational
interferometry experiment\cite{Ahmedov,Ciufolini,Hayasaka}, it becomes
possible for us to investigate quantum mechanics in weak-gravity field. The
utilization of these relativistic quantum gravitational effects enables
physicists to test the fundamental principles of general relativity in
microscopic areas. Although the equivalence principle still holds in the
relativistic quantum gravitational effect\cite{Mashhoon2}, there are some
physically interesting phenomena such as the violation of the principle of
free falling body for the spinning particle\cite{Mashhoon2,Mashhoon3} moving
in, for instance, the Kerr spacetime.

Since the analogy can be drawn between gravity and electromagnetic force in
some aspects, Aharonov and Carmi proposed the geometric effect of the vector
potential of gravity, and Anandan, Dresden and Sakurai et al. proposed the
quantum-interferometry effect associated with gravity\cite{Anandan,Dresden}.
In the rotating reference frame, a particle was acted on by the inertial
centrifugal force and Coriolis force, which are respectively analogous to
the electric force and magnetic force in electrodynamics\cite{Shen}. The
matter wave in the rotating frame propagating along a closed path will thus
possess a nonintegral phase factor (geometric phase factor), which has been
called the Aharonov-Carmi effect, or the gravitational Aharonov-Bohm effect.
Overhauser, Colella\cite{Overhauser}, Werner and Standenmann et al.\cite
{Werner} have proved the existence of the Aharonov-Carmi effect by means of
the neutron-gravity interferometry experiment. Note, Aharonov-Carmi effect
results from the interaction between the momentum of a particle and the
rotating frame. Although the interaction of a spinning particle such as
neutron with the rotating frame has the same origin of the Aharonov-Carmi
effect, i.e., both arise from the presence of the Coriolis force, the
Aharonov-Carmi effect mentioned above does not contain the spin-rotation
coupling. In the following we will propose another geometric effect that a
spinning particle possesses a geometric phase in the time-dependent rotating
frame.

Berry$^{,}$s theory of the geometric phase proposed in 1984 is applicable
only to the case of adiabatic approximation \cite{Berry}. In 1991, on the
basis of the Lewis-Riesenfeld invariant theory \cite{Lewis}, Gao et al.
proposed the invariant-related unitary transformation formulation that is
appropriate to treat the cases of non-adiabatic and non-cyclic process\cite
{Gao1}. Hence, the Lewis-Riesenfeld invariant theory is developed into a
generalized invariant theory which is a powerful tool to investigate the
geometric phase factor\cite{Gao3,Gao4}. In Sec.2, the time-dependent
spin-rotation coupling is taken into consideration by using these invariant
theories, and then we obtain exact solutions of the time-dependent
Schr\"{o}dinger equation which governs the interaction between a spinning
particle and the time-dependent rotating reference frame.

\section{Gravitomagnetic field and spin-rotation coupling}

The Kerr metric of the exterior gravitational field of the rotating
spherically symmetric body is of the form

\begin{eqnarray}
ds^{2} &=&(1-\frac{2GMr}{c^{2}(r^{2}+a^{2}\cos ^{2}\theta )})c^{2}dt^{2}-%
\frac{r^{2}+a^{2}\cos ^{2}\theta }{r^{2}+a^{2}-\frac{2GMr}{c^{2}}}dr^{2}
\nonumber \\
&&-(r^{2}+a^{2}\cos ^{2}\theta )d\theta ^{2}-\sin ^{2}\theta (\frac{%
2a^{2}\sin ^{2}\theta }{r^{2}+a^{2}\cos ^{2}\theta }\frac{GMr}{c^{2}}
\nonumber \\
&&+r^{2}+a^{2})d\varphi ^{2}+\frac{2a\sin ^{2}\theta }{r^{2}+a^{2}\cos
^{2}\theta }\frac{GMr}{c}dtd\varphi ,  \label{eq2}
\end{eqnarray}
where $r,\theta ,\varphi $ are the displacements of spherical coordinate, $a$
is so defined that $ac$ is the angular momentum of unit mass of the
gravitational body, and $M$ denotes the mass of this gravitational body.
Since the space-time coordinate of Kerr metric (\ref{eq2}) is in the fixing
reference frame, we can transform it into that in the rotating reference
frame. Because of the smallness of the Earth$^{,}$s rotating velocity, one
can apply the following Galileo transformation to the coordinates of a
particle moving radially in the rotating frame

\begin{equation}
dr^{^{\prime }}+vdt^{^{\prime }}=dr,\quad d\theta ^{^{\prime }}=d\theta
,\quad d\varphi ^{^{\prime }}=d\varphi +\omega dt,\quad dt^{^{\prime }}=dt
\label{eq3}
\end{equation}
with $v$ being the radial velocity of the particle relative to the rotating
reference frame, $($ $r^{^{\prime }},\theta ^{^{\prime }},\varphi ^{^{\prime
}},t^{^{\prime }})$ and $(r,\theta ,\varphi ,t)$ the space-time coordinates
of the rotating frame and fixing frame, respectively. $\omega $ denotes the
rotating frequency of the rotating frame with respect to the fixing
reference frame. For the simplicity of calculation , the radial velocity $v$
is taken to be much less than $\omega r,$ then substitution of Eq. (\ref{eq3}%
) into Eq. (\ref{eq2}) yields

\begin{eqnarray}
ds^{2} &=&[1-\frac{2GMr}{c^{2}(r^{2}+a^{2}\cos ^{2}\theta )}-\frac{%
(r^{2}+a^{2}\cos ^{2}\theta )}{r^{2}+a^{2}-\frac{2GMr}{c^{2}}}\frac{v^{2}}{%
c^{2}}  \nonumber \\
&&-\sin ^{2}\theta (r^{2}+a^{2}+\frac{2a^{2}\sin ^{2}\theta }{%
r^{2}+a^{2}\cos ^{2}\theta }\frac{GMr}{c^{2}})\frac{\omega ^{2}}{c^{2}}
\nonumber \\
&&-\frac{2a\sin ^{2}\theta }{r^{2}+a^{2}\cos ^{2}\theta }\frac{GMr}{c}\frac{%
\omega ^{2}}{c^{2}}]c^{2}dt^{^{\prime }2}-\frac{r^{2}+a^{2}\cos ^{2}\theta }{%
r^{2}+a^{2}-\frac{2GMr}{c^{2}}}dr^{^{\prime }2}  \nonumber \\
&&-(r^{2}+a^{2}\cos ^{2}\theta )d\theta ^{^{\prime }2}-\sin ^{2}\theta (%
\frac{2a^{2}\sin ^{2}\theta }{r^{2}+a^{2}\cos ^{2}\theta }\frac{GMr}{c^{2}}
\nonumber \\
&&+r^{2}+a^{2})d\varphi ^{^{\prime }2}-\frac{2(r^{2}+a^{2}\cos ^{2}\theta )}{%
r^{2}+a^{2}-\frac{2GMr}{c^{2}}}\frac{v}{c}dr^{^{\prime }}cdt^{^{\prime }}
\nonumber \\
&&+[\frac{2a\sin ^{2}\theta }{r^{2}+a^{2}\cos ^{2}\theta }\frac{GMr}{c}%
+2\sin ^{2}\theta (r^{2}+a^{2}+  \nonumber \\
&&\frac{2a^{2}\sin ^{2}\theta }{r^{2}+a^{2}\cos ^{2}\theta }\frac{GMr}{c^{2}}%
)\omega ]dt^{^{\prime }}d\varphi ^{^{\prime }},  \label{eq4}
\end{eqnarray}
where $\frac{\omega ^{2}r^{2}}{c^{2}}\sin ^{2}\theta $ in $g_{tt}$ results
in the inertial centrifugal force written as $\vec{F}=m\vec{\omega}\times (%
\vec{\omega}\times \vec{r}).$ Ignoring the terms associated with $\frac{a^{2}%
}{r^{2}}\ll 1$ in $g_{t\varphi ^{^{\prime }}},$ one can obtain

\begin{eqnarray}
g_{t\varphi ^{^{\prime }}}d\varphi ^{^{\prime }}dt^{^{\prime }} &=&(\frac{%
2aGMr\sin ^{2}\theta }{cr^{2}}+2\omega r^{2}\sin ^{2}\theta )dt^{^{\prime
}}d\varphi ^{^{\prime }}  \nonumber \\
&=&(\frac{2aGM\sin \theta }{cr^{2}}+2\omega r\sin \theta )dt^{^{\prime
}}r\sin \theta d\varphi ^{^{\prime }}.  \label{eq5}
\end{eqnarray}
Thus the gravitomagnetic potentials can be written as
\begin{equation}
A_{\varphi }=\frac{2aGM\sin \theta }{cr^{2}}+2\omega r\sin \theta ,\quad
A_{r}=-2v,\quad A_{\theta }=0.  \label{eq6}
\end{equation}
It follows that the first term $\frac{2aGM\sin \theta }{cr^{2}}$ of $%
A_{\varphi }$ is exactly analogous to the magnetic potential $\frac{\mu _{0}%
}{4\pi }\frac{ea}{r^{2}}\sin \theta $ of the rotating charged spherical
shell in the electrodynamics. Then we can calculate the exterior
gravitomagnetic strength of the rotating gravitational body, and the result
is $\vec{B}_{g}=\frac{2G}{c}(\frac{\vec{a}}{r^{3}}-\frac{3(\vec{a}\cdot \vec{%
r})\vec{r}}{r^{5}})\cite{Ahmedov}.$

In accordance with the equation of geodesic line of a particle in the
post-Newtonian approximation, the gravitomagnetic strength can be defined by
$-\frac{1}{2}\nabla \times \vec{A}$ with $\vec{A}=(g_{01},g_{02},g_{03})$ as
assumed above. Set \ $\beta _{\varphi }=2\omega r\sin \theta ,\beta
_{r}=-2v,\beta _{\theta }=0,$ then the gravitomagnetic strength that arises
from the choice of the reference frames is given as follows:

\begin{equation}
-\frac{1}{2}\nabla \times \vec{\beta}=-2\omega \cos \theta e_{r}+2\omega
\sin \theta e_{\theta }  \label{eq7}
\end{equation}
with $e_{r},e_{\theta }$ being the unit vector. It follows from Eq.(\ref{eq7}%
) that this gravitomagnetic strength is related to the rotation of
noninertial frame and independent of the Newtonian gravitational constant $%
G. $ From the point of view of Newtonian mechanics, it is the inertial force
field in essence rather than the field that is produced by mass current.
Since we have assumed that the velocity of a particle is parallel to $e_{r},$
i.e., $\vec{v}=ve_{r},$ the gravitational Lorentz force acting on the
particle in the gravitomagnetic field is thus given by
\begin{equation}
\vec{F}=m\vec{v}\times (-\frac{1}{2}\nabla \times \vec{\beta})=2v\omega \sin
\theta e_{\varphi }=2m\vec{v}\times \vec{\omega},  \label{eq8}
\end{equation}
We conclude from Eq. (\ref{eq8}) that the gravitational Lorentz force in
rotating reference frame is the familiar Coriolis force and the rotating
frequency $\vec{\omega}$ can be regarded as the gravitomagnetic field
strength.

In the following we will derive the Hamiltonian of spin-rotation coupling by
investigate the Dirac equation with spin connection

\begin{equation}
\lbrack i\gamma ^{\mu }(\partial _{\mu }-\frac{i}{4}\sigma ^{\lambda \tau
}\omega _{\lambda \tau \mu })-mc]\psi =0  \label{eq9}
\end{equation}
with $\sigma ^{\lambda \tau }=\frac{i}{2}(\gamma ^{\lambda }\gamma ^{\tau
}-\gamma ^{\tau }\gamma ^{\lambda }).$ In the rotating frame, we have the
following form of the line element of spacetime

\begin{equation}
ds^{2}=(1-\frac{\omega ^{2}}{c^{2}}\vec{x}\cdot \vec{x})c^{2}dt^{2}-d\vec{x}%
\cdot d\vec{x}-2(\vec{\omega}\times \vec{x})\cdot d\vec{x}dt  \label{eq11}
\end{equation}
by ignoring the gravitational effect associated with the gravitational
constant $G$ and utilizing the weak-field low-motion approximation. Then
further calculation yields the following connections\cite{Hehl}

\begin{eqnarray}
\omega _{\lambda \tau 0} &=&-\epsilon _{\lambda \tau \eta }\frac{\omega
^{\eta }}{c},\quad \omega _{0\tau 0}=-\omega _{\tau 00}=0,\quad \quad
\nonumber \\
\omega _{\lambda \tau \mu } &=&0(\mu =1,2,3)  \label{eq10}
\end{eqnarray}
with $\epsilon _{\lambda \tau \eta }$ being three-dimensional Levi-Civita
tensor. By making use of Eq. (\ref{eq9}), Eq.(\ref{eq11}) and Eq. (\ref{eq10}%
), one can arrive at the following Dirac equation

\begin{equation}
i\frac{\partial }{\partial t}\psi =H\psi  \label{eq12}
\end{equation}
with

\begin{equation}
H=\beta mc^{2}+c\vec{\alpha}\cdot \vec{p}+\vec{\omega}\cdot \vec{L}+\vec{%
\omega}\cdot \vec{S}.  \label{EQ13}
\end{equation}
We thus obtain the Hamiltonian of spin-rotation coupling

\begin{equation}
H_{s-r}=\vec{\omega}\cdot \vec{S}  \label{eq14}
\end{equation}
which is consistent with Mashhoon$^{,}$s result\cite{Mashhoon1}.

\section{Exact solutions of time-dependent spin-rotation coupling}

The variation of the Earth's rotating frequency may be caused by
the motion of interior matter, tidal force, and the motion of
atmosphere as well. Once we have information concerning the
Earth$^{,}$s rotating frequency, it is possible to investigate the
motion of matter on the Earth. For the sake of detecting the
fluctuation of the Earth$^{,}$s time-dependent rotation
conveniently, we suggest a potential approach to measuring the
geometric phase factor arising from the interaction of neutron
spin with the Earth$^{,}$s rotation by using the neutron
interferometry experiment. First we should exactly solve the
time-dependent Schr\"{o}dinger equation of a spinning particle in
the rotating system.

The Schr\"{o}dinger equation which governs the interaction of neutron spin
with Earth$^{,}$s rotation is

\begin{equation}
i\frac{\partial }{\partial t}\left| \Psi (t)\right\rangle
_{s}=H_{s-r}(t)\left| \Psi (t)\right\rangle _{s}.  \label{eq33}
\end{equation}
Set $\vec{\omega}(t)=\omega _{0}(t)[\sin \theta (t)\cos \varphi (t),\sin
\theta (t)\sin \varphi (t),\cos \theta (t)],$ and $\sigma _{\pm }=\sigma
_{1}\pm i\sigma _{2}$ with $\sigma _{1},\sigma _{2}$ being Pauli matrices$,$
then the expression (\ref{eq14}) for $H_{s-r}(t)$ can be rewritten as

\begin{eqnarray}
H_{s-r}(t) &=&\omega _{0}(t)\{\frac{1}{4}\sin \theta (t)\exp [-i\varphi
(t)]\sigma _{+}+\frac{1}{4}\sin \theta (t)\exp [i\varphi (t)]\sigma _{-}
\nonumber \\
&&+\frac{1}{2}\cos \theta (t)\sigma _{3}\}.  \label{eq341}
\end{eqnarray}

In accordance with the invariant theory, an invariant which satisfies the
following invariant equation\cite{Lewis}

\begin{equation}
\frac{\partial I(t)}{\partial t}+\frac{1}{i}[I(t),H_{s-r}(t)]=0
\label{eq340}
\end{equation}
should be constructed often in terms of the generators of Hamiltonian (\ref
{eq341}). Then it follows from Eq. (\ref{eq340}) that the invariant may be
written in terms of \ Pauli matrices as follows

\begin{equation}
I(t)=\frac{1}{4}\sin \lambda (t)\exp [-i\gamma (t)]\sigma _{+}+\frac{1}{4}%
\sin \lambda (t)\exp [i\gamma (t)]\sigma _{-}+\frac{1}{2}\cos \lambda
(t)\sigma _{3},
\end{equation}
where the time-dependent parameters $\lambda (t)$ and $\gamma (t)$ satisfy
the following two auxiliary equations

\begin{equation}
\dot{\lambda}(t)=\omega _{0}(t)\sin \theta \sin (\varphi -\gamma ),\quad
\dot{\gamma}(t)=\omega _{0}(t)[\cos \theta -\sin \theta \cot \lambda \cos
(\varphi -\gamma )]  \label{eq342}
\end{equation}
with dot denoting the time derivative. It is readily verified by using Eq. (%
\ref{eq342}) that the invariant $I(t)$ has time-independent eigenvalue $%
\sigma =\pm \frac{1}{2}$ and its eigenvalue equation is

\begin{equation}
I(t)\left| \sigma ,t\right\rangle =\sigma \left| \sigma ,t\right\rangle .
\end{equation}

According to the Lewis-Riesenfeld invariant theory, the particular solution $%
\left| \sigma ,t\right\rangle _{s}$ of Eq. (\ref{eq33}) is different from
the eigenfunction $\left| \sigma ,t\right\rangle $ of the invariant $I(t)$
only by a phase factor $\exp [i\phi _{\sigma }(t)]$. Then the general
solution of the Schr\"{o}dinger equation (\ref{eq33}) can be written as

\begin{equation}
\left| \Psi (t)\right\rangle _{s}=%
\mathop{\textstyle\sum}%
_{\sigma }C_{\sigma }\exp [i\phi _{\sigma }(t)]\left| \sigma ,t\right\rangle
,  \label{eq25}
\end{equation}
where

\begin{eqnarray}
\phi _{\sigma }(t) &=&\int_{0}^{t}\left\langle \sigma ,t^{^{\prime }}\right|
i\frac{\partial }{\partial t^{^{\prime }}}-H_{s-r}(t^{^{\prime }})\left|
\sigma ,t^{^{\prime }}\right\rangle dt^{^{\prime }},  \nonumber \\
C_{\sigma } &=&\langle \sigma ,t=0\left| \Psi (0)\right\rangle _{s}.
\label{eq26}
\end{eqnarray}

In order to obtain the analytic solution of the time-dependent
Schr\"{o}dinger equation (\ref{eq33}), we introduce an invariant-related
unitary transformation operator $V(t)$

\begin{equation}
V(t)=\exp [\frac{\beta (t)}{2}\sigma _{+}-\frac{\beta ^{\ast }(t)}{2}\sigma
_{-}],  \label{eq36}
\end{equation}
where the time-dependent parameter

\begin{equation}
\beta (t)=-\frac{\lambda (t)}{2}\exp [-i\gamma (t)],\quad \beta ^{\ast }(t)=-%
\frac{\lambda (t)}{2}\exp [i\gamma (t)].  \label{eq37}
\end{equation}
$V(t)$ can be easily shown to transform the time-dependent invariant $I(t)$
to $I_{V}(t)$ which is time-independent:

\begin{equation}
I_{V}\equiv V^{\dagger }(t)I(t)V(t)=\frac{1}{2}\sigma _{3}.  \label{eq38}
\end{equation}
The eigenstate of the $I_{V}=\frac{1}{2}\sigma _{3}$ corresponding to the
eigenvalue $\sigma $ is denoted by $\left| \sigma \right\rangle $ which is
of the form

\begin{equation}
\left| \sigma \right\rangle =V^{\dagger }(t)\left| \sigma ,t\right\rangle .
\end{equation}
By making use of $V(t)$ in expression (\ref{eq36}) and the
Baker-Campbell-Hausdorff formula\cite{Wei}, one can obtain $H_{V}(t)$ from $%
H_{s-r}(t)\cite{Gao1}$

\begin{eqnarray*}
H_{V}(t) &=&V^{\dagger }(t)H_{s-r}(t)V(t)-V^{\dagger }(t)i\frac{\partial V(t)%
}{\partial t} \\
&=&\frac{1}{2}\{[\cos \lambda \cos \theta +\sin \lambda \sin
\theta \cos (\gamma -\varphi )]+\dot{\gamma}(1-\cos \lambda
)\}\sigma _{3}.\label{eq39}
\end{eqnarray*}
From Eqs.(\ref{eq342}), it is shown that

\begin{equation}
\cos \lambda \cos \theta +\sin \lambda \sin \theta \cos (\gamma -\varphi )=0,
\label{eq312}
\end{equation}
thus, the expression (\ref{eq39}) can be rewritten as

\begin{equation}
H_{V}(t)=\frac{1}{2}\dot{\gamma}(t)[1-\cos \lambda (t)]\sigma _{3}.
\label{eq313}
\end{equation}
Based on (\ref{eq26}) and (\ref{eq313}), the geometric phase of the neutron
whose eigenvalue of spin is $\sigma $ can be expressed by

\begin{equation}
\phi _{\sigma }(t)=-\frac{1}{2}\{%
\textstyle\int%
_{0}^{t}\dot{\gamma}(t^{^{\prime }})[1-\cos \lambda (t^{^{\prime
}})]dt^{^{\prime }}\}\left\langle \sigma \right| \sigma _{3}\left| \sigma
\right\rangle .  \label{eq314}
\end{equation}

Since we know the eigenvalues and eigenstates of $I_{V}(t)=\frac{1}{2}\sigma
_{3},$ with the help of (\ref{eq25}), (\ref{eq26}), and (\ref{eq314}), it is
easy to get the general solution of the time-dependent Schr\"{o}dinger
equation which governs the neutron spin-rotation coupling is given

\begin{equation}
\left| \Psi (t)\right\rangle _{s}=%
\mathop{\textstyle\sum}%
_{\sigma }C_{\sigma }\exp [i\phi _{\sigma }(t)]V(t)\left| \sigma
\right\rangle  \label{eq315}
\end{equation}
with the coefficients $C_{\sigma }=\langle \sigma ,t=0\left| \Psi
(0)\right\rangle _{s}.$

It follows from the expression (\ref{eq312}) that the dynamical phase of
solutions of Eq. (\ref{eq33}) vanishes, and the geometric phase is expressed
by Eq. (\ref{eq314}). Since geometric phase appears only in systems whose
Hamiltonian is time-dependent or possessing some evolution parameters, this
enables us to obtain the information concerning the variation of the Earth$%
^{,}$s rotation by measuring the geometric phase difference of spin
polarized vertically down and up in the neutron-gravity interferometry
experiment.

\section{Concluding remarks}

This paper obtains the expression for the Hamiltonian of spin-rotation
coupling by coordinate transformation of Kerr metric from the fixing
reference frame to the rotating reference frame. By making use of the
Lewis-Riesenfeld invariant theory and the invariant-related unitary
transformation formulation, we obtain exact solutions of the time-dependent
Schr\"{o}dinger equation governing the interaction of neutron spin with Earth%
$^{,}$s rotation. We propose a potential method to investigate the
time-varying rotating frequency of the Earth by measuring the phase
difference between geometric phases of neutron spin down and up. In view of
the above discussions, the invariant-related unitary transformation
formulation is a useful tool for treating the geometric phase factor and the
time-dependent Schr\"{o}dinger equation. This formulation replaces the
eigenstates of the time-dependent invariants with those of the
time-independent invariants through the unitary transformation.
Additionally, it should be pointed out that the time-dependent
Schr\"{o}dinger equation is often seen in the literature, whereas the exact
solutions of time-dependent Klein-Gordon equation is paid less attention to.
Work in this direction is under consideration and will be published
elsewhere.

Acknowledgment%
%
This project was supported by the National Natural Science Foundation of
China under the project No.$30000034$.

\end{document}